\documentclass[preprint,showpacs,preprintnumbers,amsmath,amssymb]{revtex4}

\usepackage{graphicx}
\usepackage{amsmath} 

\begin{document}

\title{Local Ordering Of Polymer-Tethered Nanospheres And Nanorods And The Stabilization Of The Double Gyroid Phase}

\author{Christopher R. Iacovella$^1$}
\author{Mark A. Horsch$^1$}
\author{Sharon C. Glotzer$^{1,2}$}
\affiliation{$^1$Department of Chemical Engineering and $^2$Department of Materials Science and Engineering \\University of Michigan, Ann Arbor, Michigan 48109-2136}

\date{\today}

\begin{abstract}

We present results of Brownian dynamics simulations of tethered nanospheres and tethered nanorods.  Immiscibility between tether and nanoparticle facilitates microphase separation into the bicontinuous, double gyroid structure (first reported by Iacovella \textit{et al.} [Phys. Rev. E 75 (2007)] and Horsch \textit{et al.} [J. Chem. Phys. 125 (2006)] respectively). We demonstrate the ability of these nanoparticles to adopt distinct, minimal energy local packings, in which nanospheres form icosahedral-like clusters and nanorods form splayed hexagonal bundles.  These local structures reduce packing frustration within the nodes of the double gyroid.  We argue that the ability to locally order into stable structures is key to the formation of the double gyroid phase in these systems.\end{abstract}

\maketitle

Block copolymers and surfactants have long been known to self-assemble into a wide variety of complex structures where the assembly is driven by immiscibility between chemically distinct blocks in the polymers \cite{bates1990} and between distinct head and tail groups in surfactants \cite{larson1996}. These ordered structures are highly sought for applications at the nanoscale, ranging from photonic-bandgap materials \cite{maldovan2002} to templates for nanoparticle assembly \cite{pine2005} and hydrogen storage.   Hybrid building blocks have recently been created that resemble block copolymers where the individual blocks consist of nanoparticles and polymers \cite{kotov2002, song2003, frank2005, nie2007, devries2007}. These hybrid building blocks, or tethered nanoparticles, constitute a class of  ``shape amphiphiles'' \cite{date2003,zhang2003} where microphase separation occurs due to the immiscibility between the nanoparticle and polymeric tether  resulting in mesostructured equilibrium phases that resemble the morphologies of block copolymers and surfactants \cite{horsch2005,iacovella2005}.  

In previous work, we examined the interplay between microphase separation and nanoparticle geometry for tethered nanospheres (TNS) \cite{iacovella2007, iacovella2005} and tethered nanorods (TNR) \cite{horsch2005, horsch2006}, finding unique changes to phase behavior as compared to flexible surfactants and block copolymers.  For example, the phase behavior of the nanosphere-aggregating TNS system includes hexagonally packed cylinders, the double gyroid morphology, and perforated lamellar phases where there is a predominance towards icosahedral ordering of nanospheres \cite{iacovella2007} and a lamellar phase with HCP ordering of nanospheres \cite{iacovella2007}. The tether-aggregating TNS system does not form the double gyroid structure, as would be expected for surfactants, instead forming perforated lamella \cite{iacovella2005}.  The phase behavior of the nanorod-aggregating TNR system includes a hexagonally packed cylinder phase where the nanorods twist along the length of the cylinder, hexagonally and tetragonally perforated lamellar phases where the nanorods form a smectic structure, and a lamellar phase with smectic C-like ordering of the nanorods \cite{horsch2005}.  A transition from hexagonally packed cylinders to the double gyroid morphology is seen upon reducing the length of the tether in the TNR system \cite{horsch2006}, similar to what was observered for rod-coil liquid crystals \cite{lee1998}.

In this work, we examine and compare the double gyroid (DG) microstructure formed by the TNS and TNR building blocks with attractive nanoparticles and repulsive tethers in order to learn about the stability of the DG structure.  The DG is a bicontinous structure where the nanoparticles form two distinct, interpenetrating networks. The DG structure is of particular interest as it is seen as a candidate for catalytic materials, high conductivity nanocomposites \cite{cho2004}, and photonics applications \cite{maldovan2002}.  In this work we first present our TNS and TNR models and simulation method in section \ref{sec:modelmethod}, followed by the calculation of the Flory-Huggins interaction parameter to allow for comparisons between the TNS and TNR systems in section \ref{sec:fhTheory}.  In section \ref{sec:resultsPacking} we investigate packing frustration within the DG and its connection to particle geometry. In sections \ref{sec:resultsTNS} and \ref{sec:resultsTNR} we investigate the local configurations of the nanospheres and nanorods respectively. In section \ref{sec:conclusions} we provide concluding remarks.

\section{\label{sec:modelmethod}Model and Method}
To study tethered nanoparticles, we consider a general class of tethered nanoparticles rather than any one specific system and use empirical pair potentials that have been successful in the study of block copolymers and surfactants \cite{soddemann2001}. We utilize minimal models that capture the essential physics of the problem, specifically the geometry of the nanoparticle, immiscibility between tether and nanoparticle, and flexibility of the polymer tether. We examine tethered nanospheres and tethered nanorods in selective solvent and additionally explore the connection between these building blocks and diblock copolymers in selective solvent. The natural units of these systems are: $\sigma$, the diameter of a tether bead; \textit{m}, the mass of a tether bead; and $\epsilon$, the Lennard-Jones well depth. Bulk system volume fraction, $\phi$, is defined as the ratio of volume of the beads to the system volume, the dimensionless time is  \textit{t}*=$\sigma\sqrt{m/\epsilon}$, and the degree of immiscibility and solvent quality are determined by the inverse temperature, 1/\textit{T*} = $\epsilon/k_BT$.  

\begin{figure}[ht]
\includegraphics[width=3.25in]{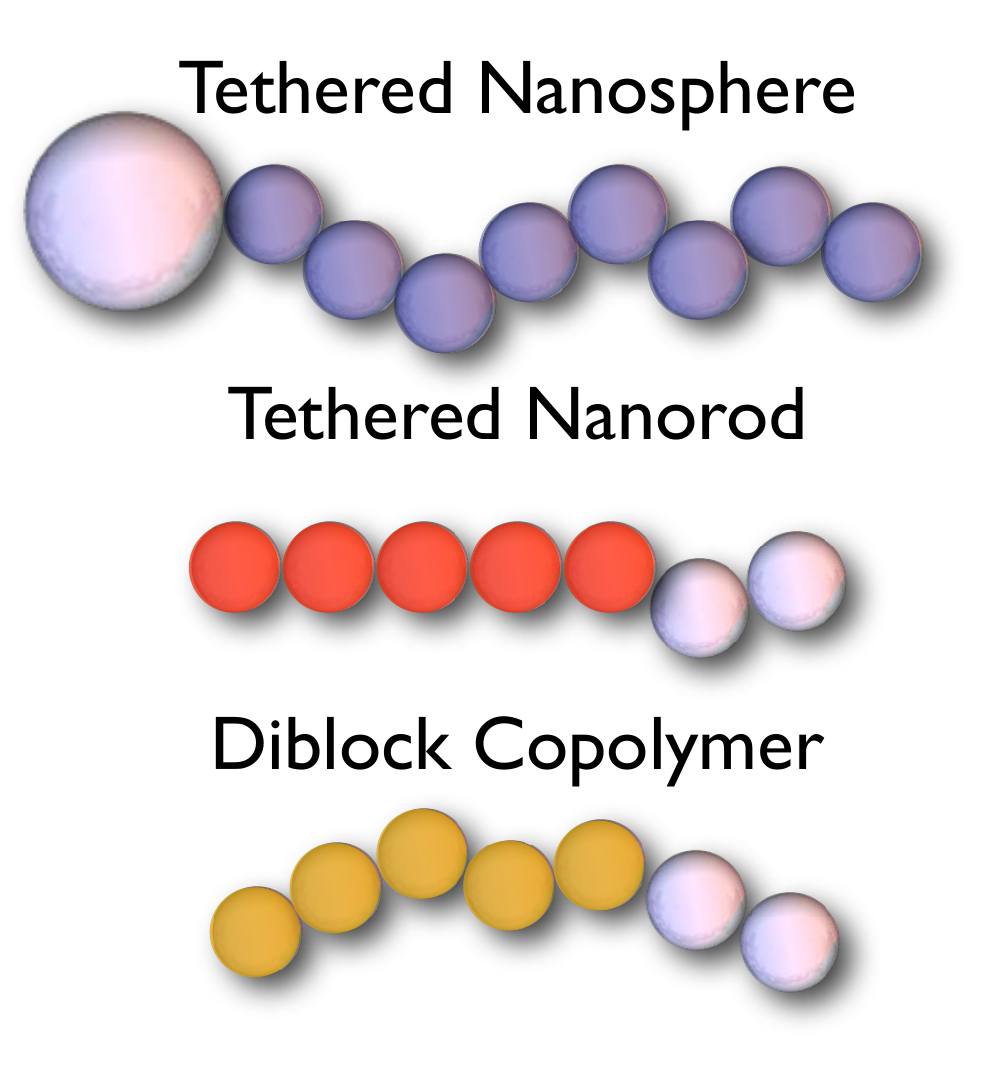} 
\caption{Model building blocks utilized.}
\label{figureModel}
\end{figure}

\subsection{\label{sec:modelTNS}Tethered Nanospheres}
Nanospheres are modeled as beads of diameter 2.0$\sigma$ connected to tethers via finitely extensible non-linear elastic (FENE) springs \cite{grest1986}. Tethers are modeled as bead-spring chains containing eight beads of diameter $\sigma$ connected via FENE springs; a schematic of the model building block is shown in Figure \ref{figureModel}. To model the attractive interaction between NPs we use the Lennard-Jones potential (LJ) where particle-particle interactions are shifted to the surface (Equation \ref{eqnLJ}); for nanoparticle-nanoparticle interactions we set $r_{shift} = (2.0\sigma-\sigma)$ and $r_{cutoff} = 2.5\sigma+r_{shift}$.
\begin{equation}
U_{LJS} = 
\begin{cases}
4 \epsilon \left( \frac{\sigma^{12}}{(r-shift)^{12}}-\frac{\sigma^{6}}{(r-shift)^6} \right) - 4 \epsilon \left( \frac{\sigma^{12}}{(2.5)^{12}}-\frac{\sigma^{6}}{(2.5)^6} \right)\;, & r<r_{cutoff}\\
0\;, &  r \ge r_{cutoff}
\end{cases}
\label{eqnLJ}
\end{equation}
Solvophilic tethers interact via the purely repulsive Weeks-Chandler-Andersen (WCA) soft-sphere potential to account for short-range, excluded volume interactions (Equation \ref{eqnWCA}); for tether-tether interactions $r_{shift} = 0$  and  $r_{cutoff} = 2^{1/6}\sigma$. 
\begin{equation}
U_{WCA} = 
\begin{cases}
4 \epsilon \left( \frac{\sigma^{12}}{(r-r_{shift})^{12}}-\frac{\sigma^{6}}{(r-r_{shift})^6} \right) + \epsilon\;, & r<r_{cutoff} \\
0\;, &  r \ge r_{cutoff}
\end{cases}
\label{eqnWCA}
\end{equation}
Nanosphere-tether interactions are treated with the purely repulsive WCA soft-sphere potential to account for short-range, excluded volume interactions (Equation \ref{eqnWCA}); for nanosphere-tether interactions $r_{shift} = \frac{1}{2}(2.0\sigma-\sigma)$  and  $r_{cutoff} = 2^{1/6}\sigma+r_{shift}$. This model relates well to experimentally synthesized building blocks including tethered quantum dots \cite{kotov2002}, polymer-functionalized fullerenes \cite{song2003},  tethered nanospheres formed by crosslinking one block of a BCP \cite{frank2005},  and divalent nanospheres \cite{devries2007}. 

\subsection{\label{sec:modelTNR}Tethered Nanorods}
Nanorods are modeled as a rigid collection of five beads of diameter $\sigma$ connected to tethers via FENE springs; a schematic of the model building block is shown in Figure \ref{figureModel}. The beads in the rod are spaced a distance $\sigma$ apart creating a relatively ``rough'' rod for computational efficiency.  In previous studies the assembly behavior of a ``rough'' rod  was indistinguishable from that of a ``smooth'' rod \cite{horsch2006}. Moreover, this rough rod relates well to colloidal rod assemblies that have recently been fabricated \cite{solomon2007, siva2007}.  Tethers are modeled as bead-spring chains containing two beads of diameter $\sigma$ connected via FENE springs.  Interactions between nanorods are modeled by the LJ potential (Equation \ref{eqnLJ}), where $r_{shift} =0$ and $r_{cutoff}=2.5\sigma$.  Solvophilic tethers and species of different type interact via the WCA potential (Equation \ref{eqnWCA}) to account for short-range, excluded volume interactions, where in all cases $r_{shift} =0$ and $r_{cutoff}=2^{1/6}\sigma$.

\subsection{\label{sec:modelBCP}Block Copolymers}
Diblock copolymer systems (BCPs) are modeled as bead spring chains where individual beads of diameter $\sigma$ are connected via FENE springs; a schematic of the model building block is shown in Figure \ref{figureModel}. Particles in the A-block  (analogous to the head group in the TNS and TNR systems) are modeled as five beads that interact via the LJ potential (Equation \ref{eqnLJ}), where $r_{shift} =0$ and $r_{cutoff}=2.5\sigma$.   Particles in the B-block (analogous to the tethers in the TNS and TNR systems) are modeled by two beads of diameter $\sigma$.   Solvophilic B-blocks (tethers) and species of different type interact via the WCA potential (Equation \ref{eqnWCA}) to account for short-range, excluded volume interactions where in all cases $r_{shift} =0$ and $r_{cutoff}=2^{1/6}\sigma$.

Additional explanation and utilization of  these models can be found in references \cite{iacovella2007, zhang2003} for TNS, \cite{horsch2006, horsch2005, zhang2003} for TNR, and \cite {iacovella2005, horsch2004} for BCP.  

\subsection{\label{sec:methodBD}Method of Brownian Dynamics}
To realize the long time scales and large systems required to study the self-assembly of complex mesophases, we use the method of Brownian dynamics (BD).  The trajectory of each ÒbeadÓ is governed by the Langevin equation:
\begin{equation}
m_{i}\mathbf{\ddot{r}}_{I}(t)= \mathbf{F}_{i}^C(\mathbf{r}_{i}(t))+\mathbf{F}_{i}^R(t) -\gamma\mathbf{v}_{i}(t)
\label{eqnBD}
\end{equation}
where $\textit{m}_i$, $\mathbf{r}_i$, $\mathbf{v}_i$, $\mathbf{F}_i^C$, $\mathbf{F}_i^R$ and $\gamma_i$ correspond to the mass, position, velocity, conservative force, random force, and friction coefficient of bead \textit{i}, respectively.  We assume that there are no spatial or temporal fluctuations in the friction coefficient and fix $\gamma_i=1.0$, which limits the range of ballistic motion of a bead to approximately 1.0$\sigma$.  The random force is independent of the conservative force and satisfies the fluctuation-dissipation theorem:
\begin{equation}
\begin{array}{l}
\langle \mathbf{F}_{i}^R(t) \rangle = 0 \\
\langle \mathbf{F}_{i}^R(t) \mathbf{F}_{j}^R(t') \rangle = 6\gamma k_{B}T\delta_{ij}\delta(t-t')
\end{array}
\end{equation}
The friction coefficient and random force act as a non-momentum-conserving heat bath; the combination of these two terms helps to minimize numerical roundoff errors that can occur over long simulations runs.  The stationary solution of the Langevin equation is the Boltzmann distribution and therefore BD samples the canonical (NVT) ensemble. We do not explicitly include solvent particles; however, the frictional and random forces help to implicitly account for some of the effects of solvent.  To incorporate rotational degrees of freedom into nanorods, we utilize the equations of rotation for linear bodies \cite{allentildesley}. In all cases, particle beads are advanced through time using the leapfrog algorithm with a timestep $\Delta t=0.01$.  The general procedure employed is to start initially at an athermal condition, where all interactions are treated using the WCA potential, and allow the system to become well mixed.   For a fixed $\phi$, we then incrementally cool the system towards a final target temperature \textit{T*} with selective solvent interactions, allowing it to run for several million timesteps at each intermediate \textit{T*}.  We also perform heating runs, where we start from the final structure and incrementally raise \textit{T*} until we reach a disordered structure.  To help avoid kinetically arrested structures, we repeat the simulations, varying the cooling sequence; for each building block, three unique cooling sequences were used.  Approximately 50 individual points were simulated for the TNS DG phase and 40 for the TNR DG phase, in total using approximately 10000 cpu hours.  Simulations were performed with 500 TNS building blocks and 800 TNR building blocks, as these correspond to the unit cell of the DG morphology.

\section{\label{sec:fhTheory}Flory-Huggins $\chi$ Parameter}
Although the models for the TNS, TNR and BCP are very similar, key parameters such as \textit{T*} do not necessarily correspond to the same statepoint.  For example, in Reference \cite{horsch2005} the rigidity of the tethered nanorod induced phase separation at a higher \textit{T*} than the equivalent coil-coil BCP.  Additionally, it is well known that the order-disorder temperature in BCP systems will scale as the number of beads \cite{groot1999}, so we expect that a tethered nanorod system with a 5-bead rod would order at a much higher \textit{T*} than a tethered nanosphere system where the nanoparticle is only a single bead, if the beads interactions were similar.  In order to compare between the TNS, TNR, and BCP systems we determine the relationship between  Flory-Huggins interaction parameter, $\chi$, and \textit{T*} for each of the systems.  The Flory-Huggins interaction parameter allows us to make comparisons between systems regardless of the specific model details used.  For example, $\chi$ has been used to compare the calculated phase boundaries of BCPs for Brownian dynamics systems that utilize attractive LJ interactions, dissipative particle dynamics systems that utilize only repulsive interactions, and mean field theory calculations \cite{horsch2004}. To determine $\chi$, we follow a similar procedure to that outlined in References \cite{groot1999} and \cite{horsch2004}.  For a two-component mixture the free energy can be expressed as:
\begin{equation}
\frac{F}{k_bT}=\frac{f_A}{v_AN_A}lnf_A+\frac{1-f_A}{v_BN_B}ln(1-f_A)+\frac{\chi f_A(1-f_A)}{(v_Av_B)^{1/2}}
\label{eqnFE}
\end{equation}
where $f_A$ and $(1-f_A)$ are the fraction of constituents A and B, respectively, $v_A$ and $v_B$ are the volume of the beads of constituents A and B, respectively, and  $N_A$ and $N_B$ are the number of beads of constituents A and B, respectively \cite{larsonbook}.  This expression assumes incompressibility where $f_B = (1-f_A)$; note $f_A$ should not be confused with the bulk volume fraction $\phi$.  If we consider the system to be in equilibrium, the free energy will be at a minimum and thus $\frac{dF}{df_A}=0$.  By taking the derivative of the free energy in Equation \ref{eqnFE} with respect to $f_A$ and setting the resulting expression equal to zero, we arrive at the following equation that relates $\chi$ to $f_A$:

\begin{equation}
\chi=-\frac{(v_Av_B)^{1/2}(-ln(f_A)v_BN_B-v_BN_B+ln(1-f_A)v_AN_A+v_AN_A)}{v_AN_Av_BN_B(2f_A-1)}
\label{eqnChi}
\end{equation}

\subsection{\label{sec:fhChi}General procedure for determining $\chi$}
To determine how $\chi$ scales with \textit{T*} for each system, we calculate the relative solubility of species A mixing into species B as a function of \textit{T*} and solve for $\chi$ from Equation \ref{eqnChi}.  For example, in the case of the tethered nanosphere system, we specify that the nanosphere is species A and the 8 bead polymer chain is species B, and utilize the following general procedure.  Note that the immiscible components (here, nanospheres and tethers) are not bonded to each other for the determination of $\chi$. 

Utilizing a long rectangular box with aspect ratio 1:1:4 (x:y:z), we place all the nanospheres on one half of the box and all polymers on the other half, creating an interface between the two immiscible species.  For a specific \textit{T*}, we run the system for approximately 10 million timesteps monitoring the potential energy to ensure we reach equilibrium.  After these 10 million equilibration timesteps we generate an average concentration profile along the long dimension of the box by collecting data over the next 10 million time steps.  To solve for $\chi$, we calculate the average fraction of nanoparticles that have mixed into the polymer region (i.e. $f_A$) from the concentration profile, then plug this value into Equation \ref{eqnChi}, solving for $\chi$.  We then repeat for various values of \textit{T*} creating a relationship between \textit{T*} and $\chi$, as shown in Figure \ref{figureChiMapping}.  Typically, for incompressible mixtures, as \textit{T*} is decreased $\chi$ will increase and this relationship is often described as $\chi = \frac{C}{T^*}+D$ \cite{larsonbook}, where $C$ and $D$ are system dependent fitting parameters; consequently, we fit our data using linear regression to determine the relationship.  Since our systems are compressible the $\chi$ mapping depends on the bulk volume fraction, $\phi$.  Specifically, the slope of the fitting will increase as $\phi$ is increased, i.e. there will be less desire for the two systems to mix.  

\subsection{\label{sec:fhChiMapping}$\chi$ mappings for TNS, TNR, and BCP}

\begin{figure}[ht]
\includegraphics[width=5.5in]{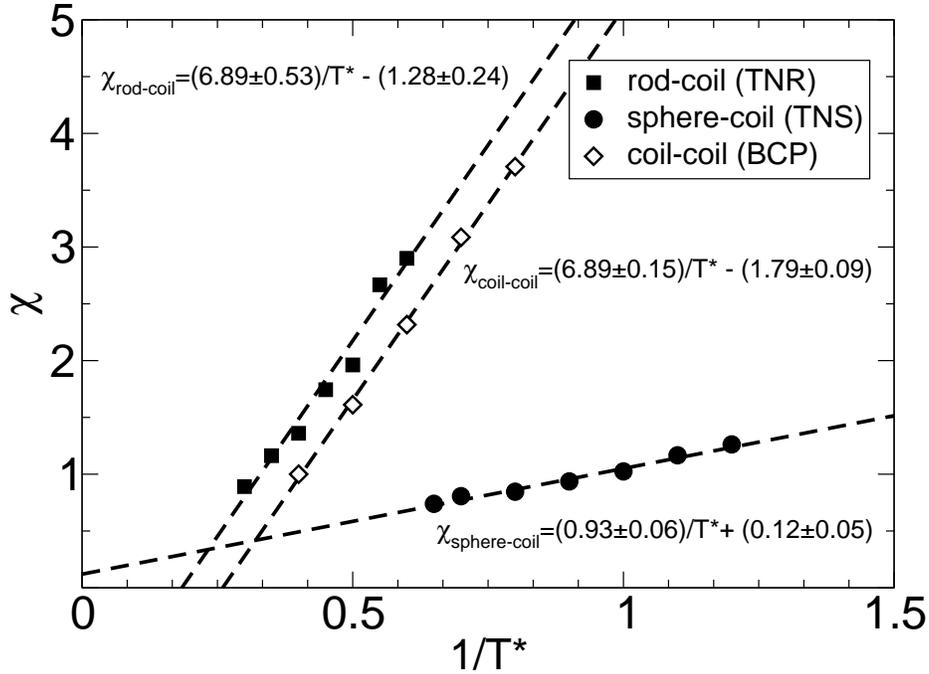} 
\caption{$\chi$ mappings for TNS (circles), TNR (squares), and BCP (diamonds).}
\label{figureChiMapping}
\end{figure}

Utilizing the same procedure for each of the building blocks, we find the following relationships for $\chi$ vs. \textit{T*} (also shown in Figure \ref{figureChiMapping}).  These mappings were performed at $\phi_{TNS}=0.3$ and $\phi_{TNR}=0.21$, corresponding to the bulk volume fractions where the DG was simulated for each of the systems, respectively; the mapping for the BCPs was performed at $\phi_{BCP}=0.21$ for appropriate comparison with the TNR system.

\begin{equation}
\chi_{TNS}=(0.93 \pm 0.06)/T^* - (0.12 \pm 0.05)
\label{eqnTNS}
\end{equation}

\begin{equation}
\chi_{TNR}=(6.89 \pm 0.53)/T^* - (1.28 \pm 0.24)
\label{eqnTNR}
\end{equation}

\begin{equation}
\chi_{BCP}=(6.89 \pm 0.15)/T^* - (1.70 \pm 0.09)
\label{eqnBCP}
\end{equation}

\noindent
These relationships follow the expected trends.  For an equivalent $\chi$ value, \textit{T*} is higher for the TNR system than the BCP system; this corresponds to the behavior seen in reference \cite{horsch2005} where the TNR system microphase separated at a higher \textit{T*} than the equivalent BCP system.   For an equivalent $\chi$ value, \textit{T*} is substantially higher for the TNR system as compared to the TNS system; previous simulations of TNRs found the system microphase separates at approximately \textit{T*}=1 \cite{horsch2005} whereas the TNS system was found to order at about \textit{T*}=0.3 \cite{iacovella2007}.  We should be aware that these results are for moderate bulk volume fractions ($\phi=0.3$ for TNS and $\phi=0.21$ for TNR and BCP); the results in References \cite{groot1999} and \cite{horsch2004} show good agreement with theory using this procedure, however the comparisons were made at higher values of $\phi$ corresponding to melt conditions ($\phi = 0.45$). The difference in $\phi$ will manifest itself in the compressibility assumption, specifically $f_B = (1-f_A)$ may not be completely valid at the lower $\phi$ values we utilize. However, since our simulations are for building blocks in selective solvent (i.e. tether beads are treated with the repulsive WCA potential), the tethers can fill the available space without forming low density holes and thus the $\chi$ mapping should allow for reasonable comparisons among these systems at the lower values of $\phi$ .

\section{\label{sec:results}Results}

\begin{figure}[ht]
\includegraphics[width=6.5in]{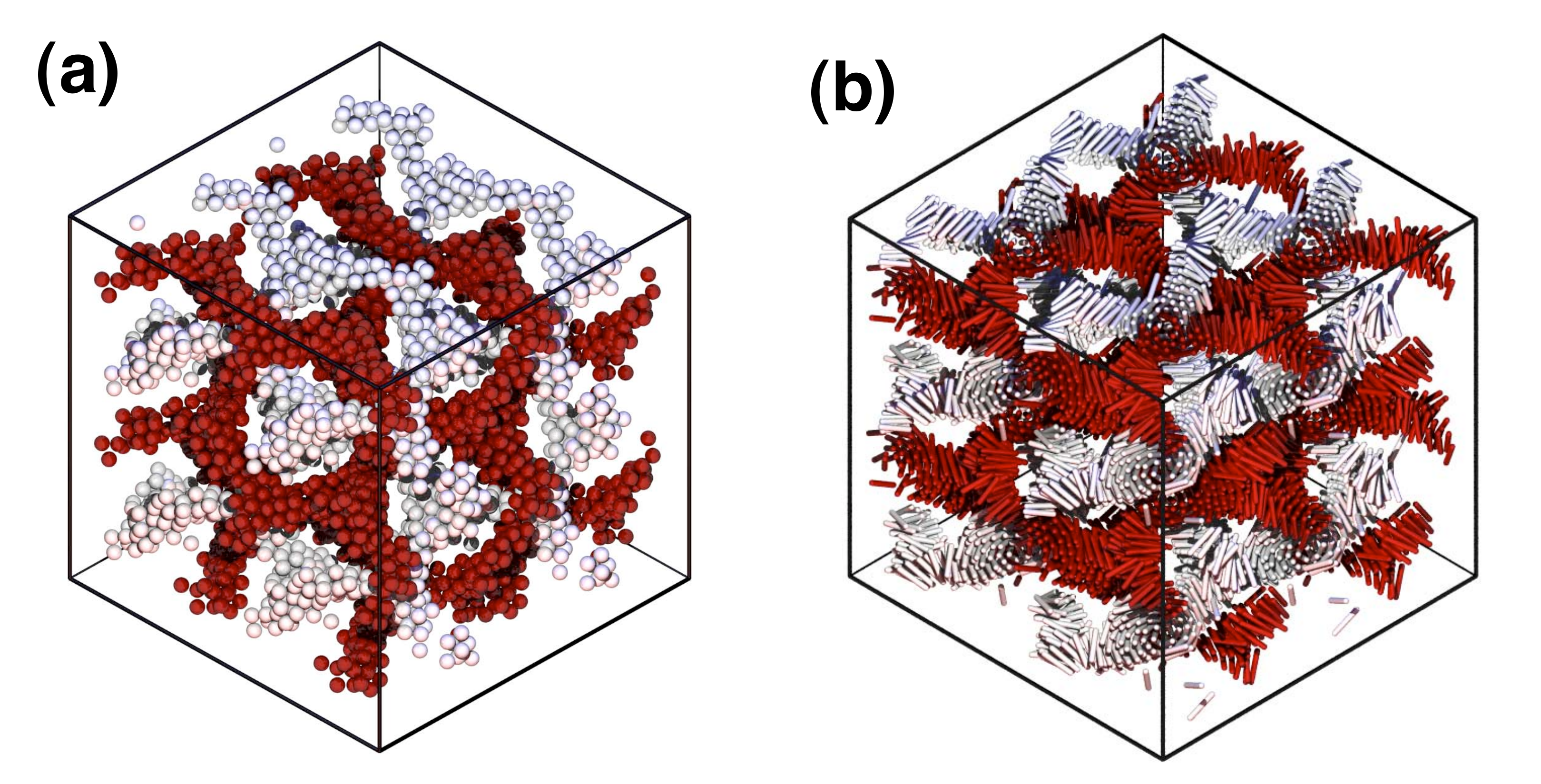} 
\caption{\textbf{(a)} DG phase formed by the TNS system, \textbf{(b)} DG phase formed by the TNR system.  In both cases tethers have been removed for clarity.  The images correspond to the minimal unit cell duplicated once in each direction for clarity.}
\label{figureStructures}
\end{figure}

In previous publications we reported the presence of the double gyroid (DG) structure in both the TNS \cite{iacovella2007} and TNR \cite{horsch2006} systems.  The double gyroid is a bicontinous structure where the minority component, in our case the nanoparticles, forms two distinct, interpenetrating networks that never connect.  The minority component organizes into a series of cylindrical tubes (arms) where three tubes connect at each node.  Figure \ref{figureStructures} shows a simulation snapshot of the DG formed by the TNS system \textbf{(a)} and the TNR system \textbf{(b)}. For clarity we have removed the tethers and show only the minority component (nanospheres or nanorods) where the two distinct networks are colored red and white; the particles in the red domain are chemically identically to those in the white domain.  These structures were identified both visually and by calculating the structure factor.  The structure factor for the DG shows two strong peaks with a characteristic ratio of $\sqrt{3}$:$\sqrt{4}$ as expected \cite{hajduk1994}.  Averaging over several runs, the DG was found to form for $\phi=0.3$ at 1/\textit{T*} $\ge 3.28$ for the TNS system and for $\phi=0.21$ at 1/\textit{T*} $\ge 0.825$ for the TNR system.  Recent theoretical predictions of the order-disorder transition for TNS agree with our simulations, reporting a value of  1/\textit{T*} $\sim$  3 for $\phi=0.3$ \cite{arthi2008}.  Using the $\chi$ mappings we calculated in section \ref{sec:fhChiMapping}, we find reasonable agreement between the average order-disorder transitions for both systems as expected.  Specifically, we find that  $\chi_{order-disorder}^{TNS} = 3.17\pm0.25$ and $\chi_{order-disorder}^{TNR} = 4.40\pm0.70$, where the reported error for $\chi$ is calculated from the maximum error in the linear regression of $\chi$ vs. \textit{T*}.  

\subsection{\label{sec:resultsPacking}Packing frustration analysis}

The formation of the DG in the tethered nanoparticle systems is surprising since it is known to exist only in a small region of the phase diagram for BCPs \cite{bates1996, escobedo2006, rychkov2005, cochran2006}, rod-coil BCPs \cite{lee1998, lee2001}, and surfactants \cite{larson1996}, and was not seen in some of our previous simulations of the TNS \cite{iacovella2005} and TNR \cite{horsch2005} systems.  The limited range of stability of the DG phase in BCP systems has been attributed to packing frustration at the nodes of the gyroid \cite{matsen1996,hasegawa1996}.  It has been shown that the standard deviation in mean curvature, $\sigma_H$, correlates to packing frustration and dictates the overall stability of a structure \cite{matsen1996}.  For example, Matsen and Bates \cite{matsen1996} calculated $\sigma_H$ for various structures finding $\sigma_H=0.121$ for the gyroid as compared to $\sigma_H=0.003$ for cylinders and attributed this difference to the inability of the gyroid structure to simultaneously minimize surface area and minimize packing frustration \cite{matsen1996}.  The magnitude of $\sigma_H$ is important in determining which phase will form as the system should selectively prefer to form a structure with less packing frustration \cite{matsen1996}; this is why BCPs typically form the DG over, for instance, the bicontinuous double diamond structure where $\sigma_H$ = 0.311  \cite{matsen1996}.  In BCPs, packing frustration has been shown to manifest itself as a high void fraction (low packing density) within the nodes of the gyroid \cite{escobedo2005} where the polymer needs to stretch to fill the volume dictated by the interface \cite{matsen1996}.   Martinez-Veracoechea and Escobedo have shown that packing frustration in the nodes of BCPs can be reduced by adding monomer and homopolymer to the system, increasing the stability of phases like the DG \cite{escobedo2005} and stabilizing other structures such as double diamond \cite{escobedo2007}.

\begin{figure}[h]
\includegraphics[width=5.5in]{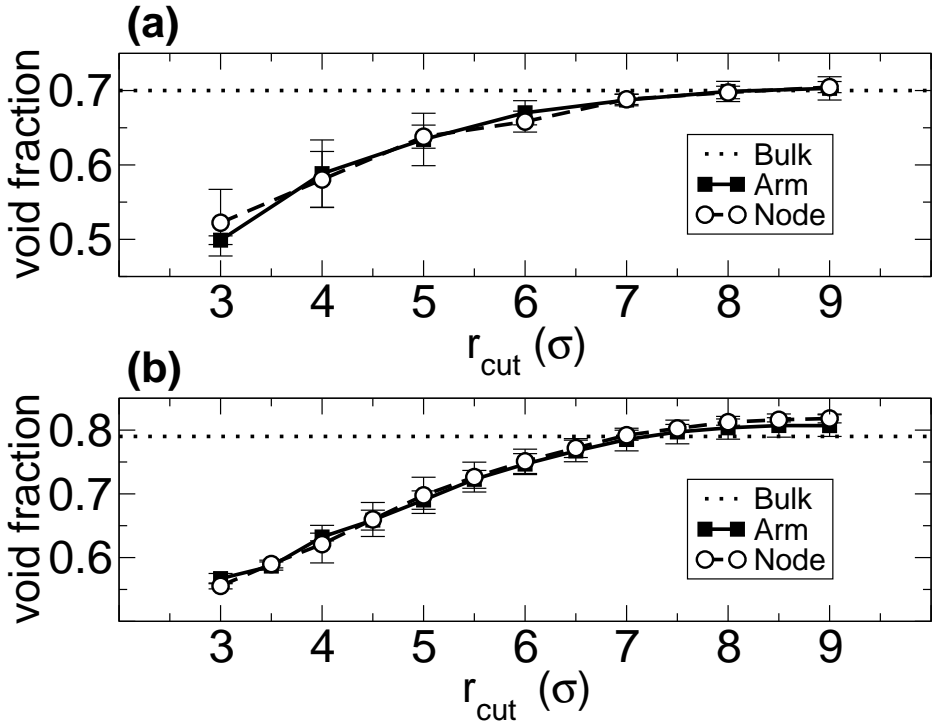} 
\caption{Relative void fraction within the nodes of the \textbf{(a)} TNS DG  for  $1/T^*$ =3.3 ($\chi$ = 3.19 $\pm$ 0.25) and \textbf{(b)} TNR DG  for $1/T^*$ =0.9 ($\chi$ = 4.92 $\pm$ 0.72). }
\label{figureNodeVoidFraction}
\end{figure}

To assess the stability of the DG structure for the TNS and TNR systems, we examine packing frustration by calculating the relative trends in void fraction at the nodes and comparing this with the void fraction of the arms, since measuring $\sigma_H$ is problematic in a system of discrete particles.  To look at this trend we approximate the center of a node and calculate the void fraction within a spherical volume drawn from the center, repeating for various sphere radii, r$_{cut}$.  If we consider a very large value of r$_{cut}$, we capture the bulk void fraction of the system and as r$_{cut}$ is decreased, we get an increasingly localized picture of what  occurs at the nodes.  We perform the same procedure on the arms, again, capturing a relative trend in void fraction.   For both the TNS and TNR systems, the void fraction of the arms and nodes are nearly identical, as shown in Figure \ref{figureNodeVoidFraction}, suggesting there is a uniform density throughout the DG in both structures, irrespective of whether we are at the node.  Hence, we do not have a characteristic high void fraction within the nodes, as has been shown for BCPs \cite{escobedo2005}, thus the nanospheres and nanorods have reduced the packing frustration as compared to a flexible BCP.   Additionally, we can compare the void fraction between the nodes and the bulk system.  For both the TNS and TNR systems, we find that for large values of r$_{cut}$ the void fraction is approximately that of the bulk system and as r$_{cut}$ is decreased, the void fraction is also decreased, as shown in Figure \ref{figureNodeVoidFraction}.  Specifically, for the TNS system, the void fraction for large values of r$_{cut}$ is $\sim$0.7 and for small values $\sim$0.5.  For the TNR system, the void fraction for large values of r$_{cut}$ is $\sim$0.79 and for small values $\sim$0.55. Thus particles within the nodes pack more densely than the bulk system as a whole.  Martinez-Veracoechea and Escobedo  used a similar analysis to compare a monodisperse BCP system to a blend of two different length BCPs \cite{escobedo2005}.   In the blend, the authors found that the longer of the two polymers occupied the nodes of the DG, resulting in a larger range of stability compared to the monodisperse system \cite{escobedo2005}.  For the monodisperse system, the authors found that the void fraction is \textit{higher} than the bulk for small values of r$_{cut}$ and for the blend the authors found that, like our systems, the void fraction is \textit{lower} than the bulk for small values of r$_{cut}$ \cite{escobedo2005}.   This similarity in trends suggests that the DG structures formed by the TNS and TNR systems may be more stable than an equivalent flexible BCP system.  We conclude that by not forming low density regions at the nodes of the DG both the TNS and TNR systems have reduced their packing frustration, and we hypothesize that this reduction is a direct result of the geometry of the minority component and its ability to pack locally into compact, low energy structures.  We test this hypothesis in sections \ref{sec:resultsTNS} and \ref{sec:resultsTNR}.   

\begin{figure}[ht]
\includegraphics[width=3.30in]{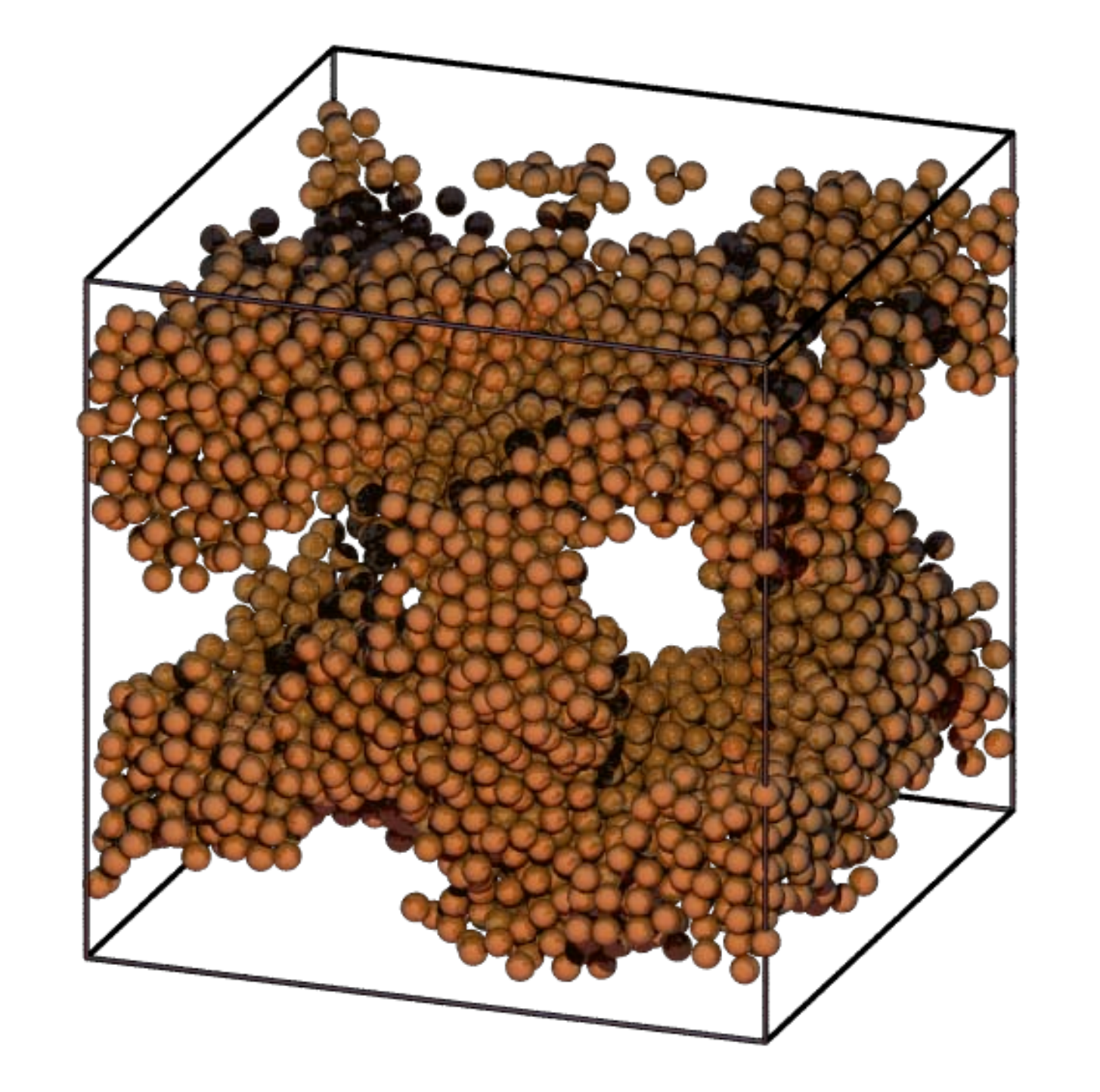} 
\caption{Disordered structure formed by BCP system for $\chi=8.6 \pm 0.26$.  The tethers have been removed for clarity, and only the minority, aggregating species is present.}
\label{figureBCPMush}
\end{figure}

We can further assess the importance of the particle geometry on stability by specifically comparing to a BCP system.  If we replace the rigid constraints in the TNR system with FENE springs, we arrive at a simple model for a flexible coil-coil BCP, as was previously described in section \ref{sec:modelBCP}.  If the rigidity of the nanorod does not influence the stability and packing frustration of the DG phase, we would expect to find the DG in the BCP system at $\chi \ge 4.40 \pm 0.70$ and $\phi=0.21$.  Starting from the ordered DG configuration, we run the system as a coil-coil BCP for various values of $\chi$.  We find that the DG structure is not stable for values of $\chi < \sim 8$; instead, the DG falls apart and forms a disordered aggregate. An example of the disordered structure is shown in Figure \ref{figureBCPMush}.  For values of $\chi > \sim 8$, the DG structure persists, however, it is most likely a kinetically arrested structure and not at equilibrium due to the large value of $\chi$.  Starting from an athermal, disordered configuration, we incrementally cool the system, finding only disordered structures, even for values of $\chi > \sim 8$.  This supports the contention that the DG is not stable for the BCP system under these conditions.  Note that simulations were also performed with various unit cell sizes to avoid any box size issues that are associated with 3d periodic microstructures, again, yielding only disordered structures.  Thus, we find that for equivalent statepoints, we were unable to realize the DG phase when the rigid constraint was removed, which suggests that the stabilization of the DG is strongly influenced and controlled by the geometry of the aggregating species. 

\subsection{\label{sec:resultsTNS}Local structure of the TNS gyroid}
\begin{figure}[ht]
\includegraphics[width=2.5in]{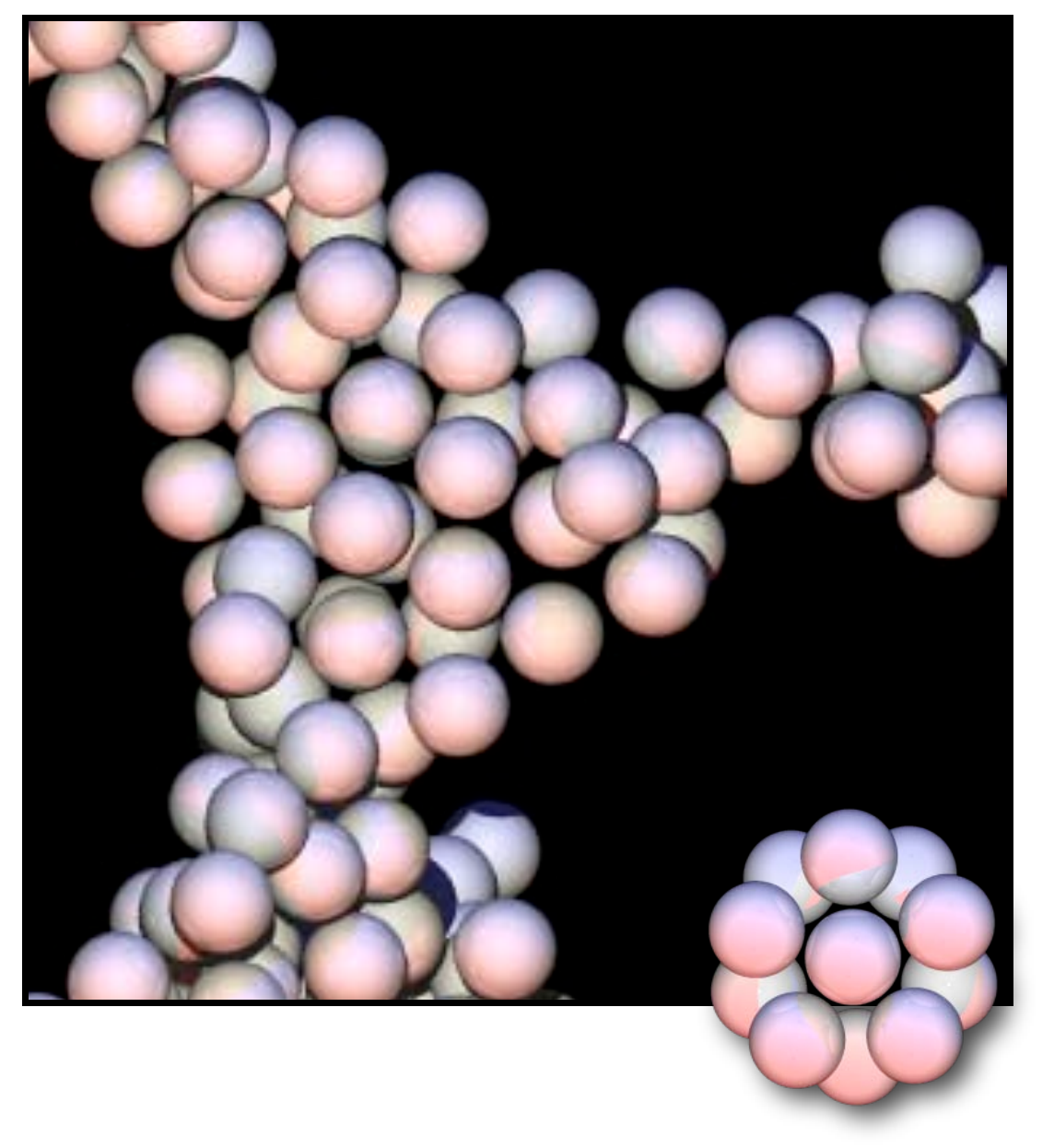} 
\caption{Node of the DG formed by the TNS system, with a perfect icosahedron showed in inset. Tethers have been removed for clarity.}
\label{figureTNSNode}
\end{figure}

The ability of the TNS system to reduce the packing frustration in the DG can be understood by looking at the local packing of the particles at the node. Figure \ref{figureTNSNode} shows a simulation snapshot of the TNS DG node.   We see distinct local ordering in the node, specifically ring-like structures that resemble icosahedral clusters.  An icosahedron is constructed of a central particle surrounded by 12 nearest neighbors and is a minimal potential energy structure for 13 Lennard-Jones particles.  This ordering is not limited to the nodes and occurs throughout the entire DG structure. 

In reference \cite{iacovella2007} we introduced the R$_{ylm}$ method based on spherical harmonics \cite{steinhardt1983} and used this to identify the local packing of these particles.  The general R$_{ylm}$ method relies on a rotationally invariant spherical harmonic ``fingerprint'' of the structure.  In this method we first calculate $q_\ell^m$ by summing over all nearest neighbor directions between particles:

\bigskip
$q_\ell^m = \frac{1}{N_b}\sum_{N_b}^{j=1}Y_{\ell m}\left[\theta(\hat{r}_{i,j}),\phi(\hat{r}_{i,j})\right]$, 
\bigskip

\noindent
where $\hat{r}_{i,j}$ is the vector drawn from particle $i$ to its nearest neighbor $j$, $N_b$ is the total number of neighbors, $\ell$ is the specific harmonic, and $Y_{\ell m}$ is the spherical harmonic expansion \cite{steinhardt1983}. From this we can construct two rotationally invariant measures of cluster shape, $Q_\ell$, the vector magnitude of  $q_\ell^m$, and $w_\ell$, the rotationally invariant combination of the average values of $q_\ell^m$ \cite{steinhardt1983}.  As defined by Steinhardt, \textit{et al.} :

\bigskip
$Q_\ell = \left((4\pi/(2\ell+1))\sum_{m=-\ell}^\ell |q_\ell^m(i)|^2\right)^{1/2}$,

\bigskip
$
w_\ell = 
\cfrac{
\displaystyle\sum_{\substack{m_1, m_2, m_3\\m_1+m_2+m_3=0}}
\begin{pmatrix}
  \ell &  \ell & \ell\\
  m_1 & m_2 & m_3
\end{pmatrix}
q_\ell^{m_1}(i)q_\ell^{m_3}(i)q_\ell^{m_3}(i)
}
{
\left(\displaystyle\sum_{m=-\ell}^\ell |q_\ell^m(i)|^2\right)^{3/2}
}
$,

\bigskip
\noindent
where $\begin{pmatrix}
  \ell &  \ell & \ell\\
  m_1 & m_2 & m_3
\end{pmatrix}$
is the Wigner 3j symbol\cite{steinhardt1983}.\bigskip

To determine the local structure of a particle we calculate the $Q_\ell$ and $w_\ell$ values for \textit{l} = 4, 6,...12, and then calculate the residual value with respect to a library of known structures with matching coordination \cite{iacovella2007}:

\bigskip
$R_i=\sqrt{\sum^{12}_{l=4}\left(Q_l-Q_{ref_i}\right)^2+\sum^{12}_{l=4}\left(w_l-w_{ref_i}\right)^2}$.
\bigskip

\noindent A particle is considered to be in the local configuration \textit{i} that minimizes the residual $R_i$ or considered to be disordered if the residual value exceeds a certain cutoff, chosen as 0.316 for this specific system and conditions.  The R$_{ylm}$ method is well suited to accurately identify particle configurations as it provides a rotationally invariant description of a local structure and does not rely on multiple cutoff values to determine the configuration, minimizing potential error.  Additional description of the method can be found in references \cite{iacovella2007, keys2008}.

\begin{figure}[h]
\includegraphics[width=3.5in]{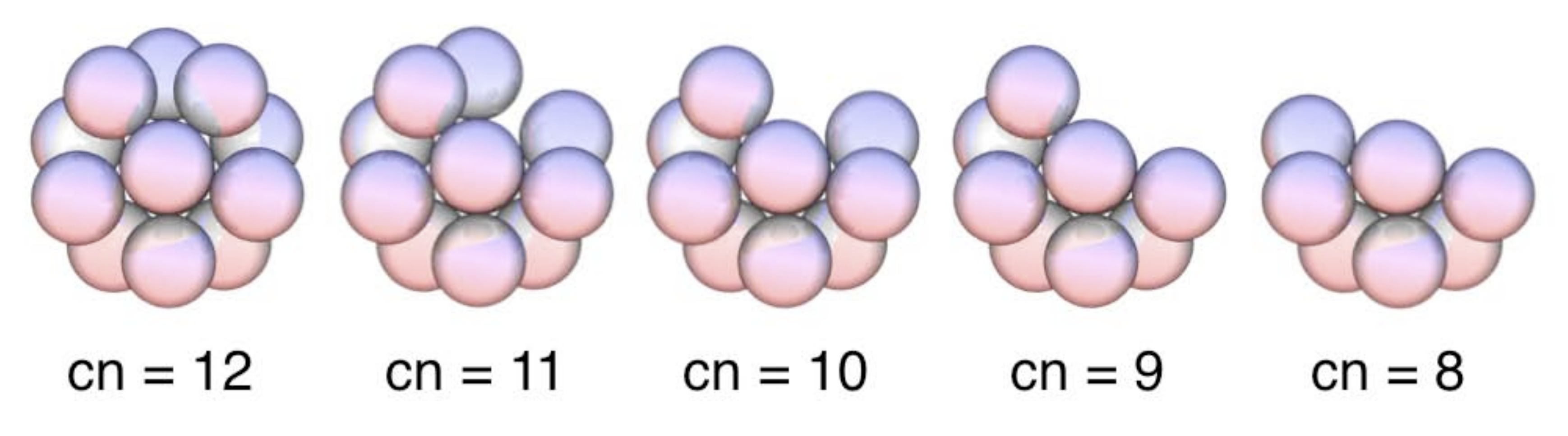} 
\caption{Icosahedral clusters ranging from full coordination number (cn) of 12 to partial coordination of 8.}
\label{figurePartialIcos}
\end{figure}

Our reference library includes standard particle arrangements such as the Kasper polyhedra (Z10, Z11, Z12, Z13, Z14, Z15) \cite{doye1996}, face centered cubic (FCC), hexagonally close packed (HCP), body centered cubic (BCC), and simple cubic (SC).  Additionally, our library includes partial icosahedral clusters, where 1-4 particles are removed from the ideal Z12 Icosahedron (see Figure \ref{figurePartialIcos}), and partial clusters of FCC and HCP where 1-5 particles are removed from the ideal clusters.

Using the R$_{ylm}$ method, we determined in reference \cite{iacovella2007} the packing of the nanoparticles in the DG to be predominantly icosahedral clusters with partial coordination, i.e. we found clusters that retain the same bond angles as a perfect icosahedron but with 1-4 particles removed (see Figure \ref{figurePartialIcos}) \cite{iacovella2007}; these clusters are identical to the minimal potential energy clusters found by Doye and Wales \cite{doye2001}. Partial clusters are formed as a result of the steric effects of the tethers and the microphase separation that occurs. Table \ref{tableIcos} shows the percentage of nanospheres that are ``central'' particles within partial icosahedral clusters, the percentage of nanospheres that are ``central'' particles within partial crystalline cluster (HCP and FCC in this case) along with the average coordination number of nanoparticles for an example cooling sequence.  The double line in Table \ref{tableIcos} signifies the transition from a disordered state to the ordered DG microphase. As we increase 1/\textit{T*} (i.e. cool the system), we notice a substantial increase in the number of icosahedral-like clusters, increasing from approximately 17.5\% to 30\% but very little change in crystalline arrangements, with an average value approximately 16\% within the ordered regime (1/\textit{T*} $>$ 3.25). We also see a minor increase in average coordination number, increasing from approximately 7 to 8; this does not change as rapidly as the percentage of icosahedral clusters since coordination number does not differentiate between ordered and disordered local configurations.

\begin{figure}[h]
\includegraphics[width=5.5in]{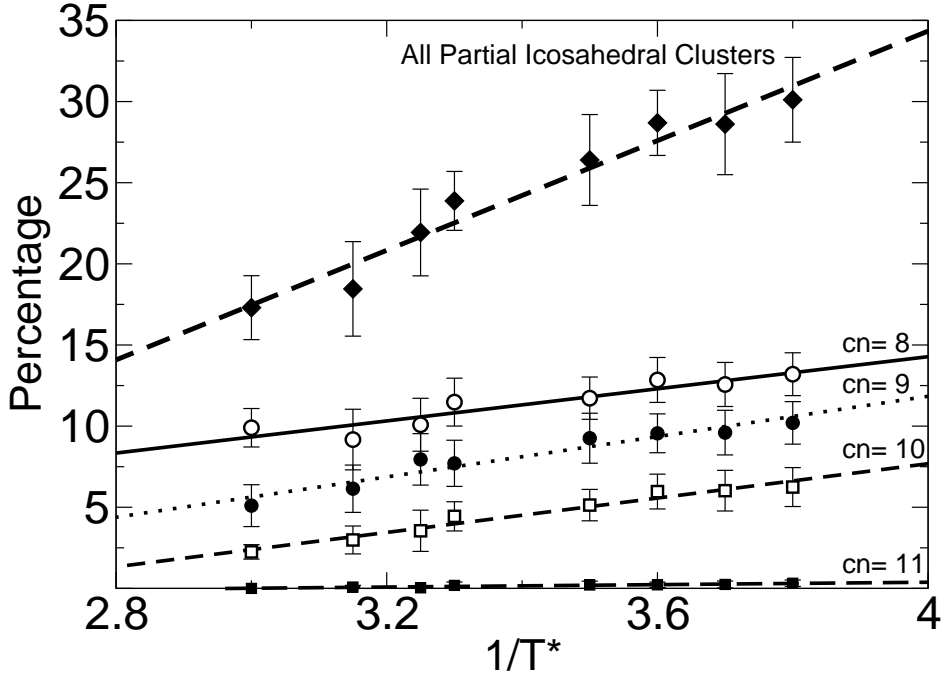} 
\caption{Icosahedral clusters with partial coordination.  Data is grouped by the coordination number of the cluster, ranging from 8 to 11.  The solid diamonds correspond to the sum of all partial icosahedral coordinations, also reported in Table \ref{tableIcos}.  All data was fit using a linear regression.  The error bars correspond to the standard deviation.}
\label{figurePartialStructures}
\end{figure}

\begin{table}[h]
\caption{Local Structure of Nanospheres}
\begin{tabular}{ccccc}
\hline
\;\; $1/T$* \;\; & \;\; $\chi$ $\pm$ error\;\; & \;\; \% Icos. $\pm$ stdev\;\; & \;\; \% Crystal $\pm$ stdev\;\; & \;\; Coord. Num. $\pm$ stdev\;\; \\
\hline
3.00 & 2.91 $\pm$ 0.23 &	17.3 $\pm$ 1.97 &	17.1 $\pm$ 1.76 & 6.9 $\pm$ 2.3\\
3.15 & 3.05 $\pm$ 0.24 &	18.5	$\pm$ 2.91 &	18.1 $\pm$  2.53 & 7.2 $\pm$ 2.1	\\
3.25 & 3.14 $\pm$ 0.25 & 	21.7	$\pm$ 2.67 &	18.7 $\pm$  2.66 & 7.4 $\pm$ 2.0	\\
\hline\hline
3.30	& 3.19 $\pm$ 0.25 & 	23.9	$\pm$ 1.81 &	16.9 $\pm$  1.91 & 7.4 $\pm$ 2.1	\\
3.50	& 3.38 $\pm$ 0.26 & 	26.4	$\pm$ 2.80 &	16.2 $\pm$  3.13 & 7.5 $\pm$ 2.0	\\
3.60	& 3.47 $\pm$ 0.27 & 	28.7	$\pm$ 2.01 &	15.7 $\pm$  2.49 & 7.6 $\pm$ 2.0	\\
3.70	& 3.56 $\pm$ 0.27 & 	28.6	$\pm$ 3.12 &	15.8 $\pm$  2.34 & 7.6 $\pm$ 2.1	\\
3.80	& 3.66 $\pm$ 0.28 & 	30.1	$\pm$ 2.61 &	15.5 $\pm$  1.62 & 7.8 $\pm$ 2.0 \\
\hline
\end{tabular}
\label{tableIcos}
\end{table}
In Figure \ref{figurePartialStructures} we also present these results grouped by the coordination of the partial icosahedral cluster, i.e. we plot the data for each cluster arrangement separately.  As the coordination number suggests, at all \textit{T*} we are most likely to find clusters with a coordination number of 8 and likewise, we find only a very small number of particles with a coordination number of 11.  The summation of all of the icosahedral clusters, as reported in Table \ref{tableIcos}, is shown as solid diamonds in Figure \ref{figurePartialStructures} demonstrating a clear linear increase over the range considered.

\begin{figure}[h]
\includegraphics[width=5.5in]{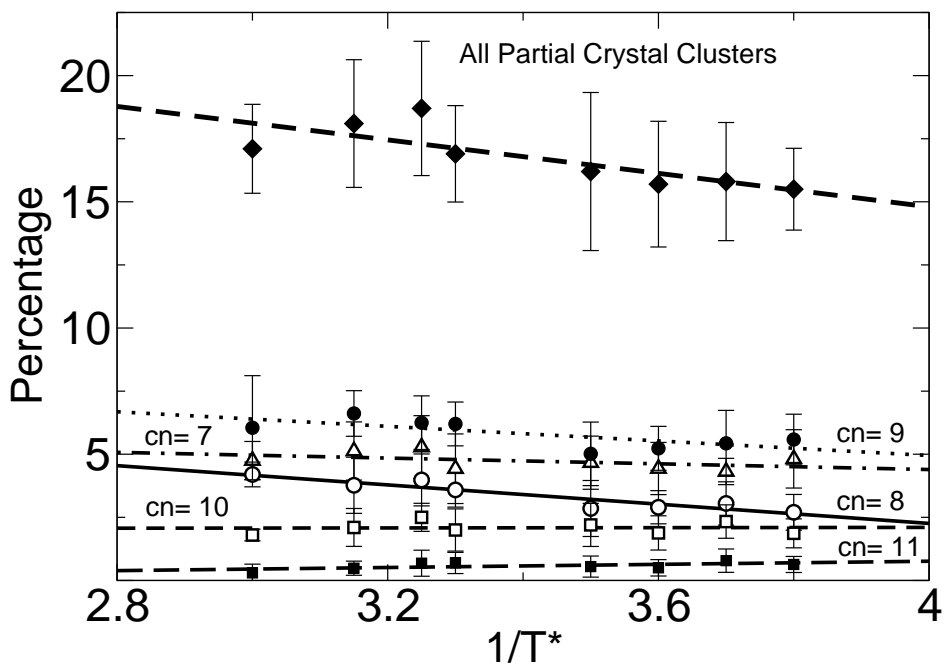} 
\caption{Percentage of nanospheres that are central particles in FCC and HCP clusters with partial coordination.  Data is grouped by the coordination number of the cluster, ranging from 7 to 11. The solid diamonds correspond to the sum of all partial crystal coordinations, also reported in \ref{tableIcos}.  All data was fit using linear regression.  The error bars correspond to the standard deviation.}
\label{figurePartialCrystal}
\end{figure}

Similarly, we can group the results for crystalline partial clusters by their coordination number.  This eliminates some complexity, since for HCP and FCC there are 12 unique partial clusters for coordination numbers ranging from 7-11; this is due to the fact that the partial cluster we form depends on which particle(s) we choose to remove.  For example, we have two unique spherical harmonic fingerprints for an HCP cluster with coordination number of 10 depending on which particles we remove.  Additionally, because of the similarities between HCP and FCC, certain partial clusters are identical and we cannot determine whether the cluster is FCC or HCP, only that it possesses characteristics of both.  As such, we group together the partial clusters of both FCC and HCP by coordination number of the partial cluster, as shown in Figure \ref{figurePartialCrystal}.  We find that for all configurations, the percentage of crystalline clusters does not change much as we increase  $1/T^*$ (i.e. cool the system); there is a minor decrease in the percentage of clusters on the order of the standard deviations.  We also see that we are most likely to have crystalline partial clusters with coordination numbers of 9 and 7; as we saw for icosahedral clusters, we find only a small amount of clusters with large coordination numbers of 11.

The fact that icosahedral arrangements are favored over crystalline ordering is somewhat surprising, as we find for a large bulk system of nanopsheres without tethers that FCC/HCP crystalline ordering is dominant; icosahedral arrangements are favored only for simulations of small numbers of particles.  In reference \cite{iacovella2007} we found that icosahedral ordering of nanoparticles was a result of the confinement that occurs as the system phase separates.  We showed that by confining particles into cylinders with hard walls, there was a transition from predominantly crystalline ordering to predominantly icosahedral ordering when the diameter of the cylinder (scaled by particle size) was less than 5, corresponding to the approximate diameter of the domains in the DG system \cite{iacovella2007}.

\subsection{\label{sec:resultsTNR}Local structure of the TNR gyroid}

\begin{figure}[ht]
\includegraphics[width=2.5in]{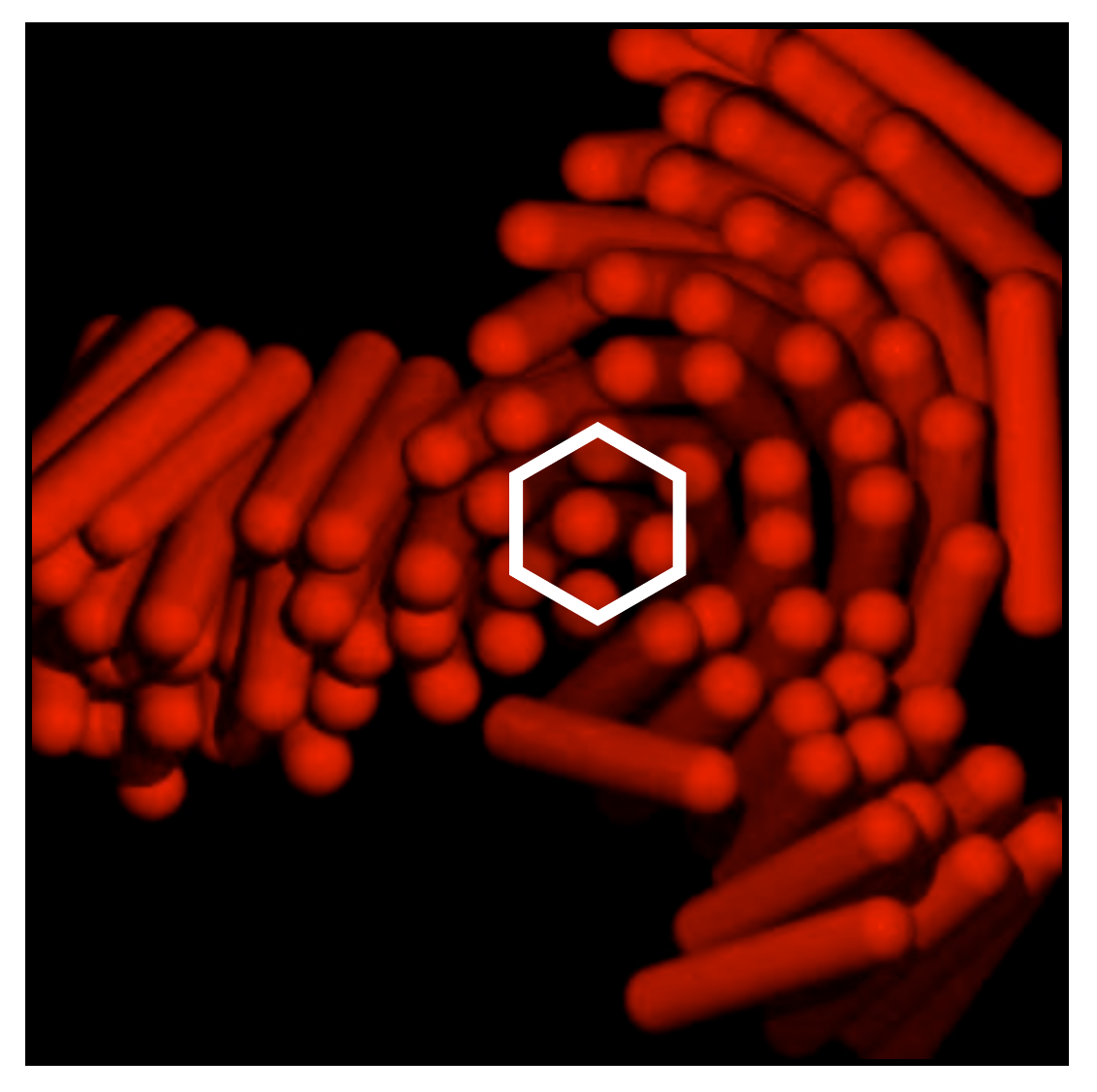} 
\caption{Node of the DG formed by the TNR system with a hexagonal bundle highlighted.  Tethers have been removed for clarity.}
\label{figureTNRNode}
\end{figure}

\begin{figure}[h]
\includegraphics[width=5.5in]{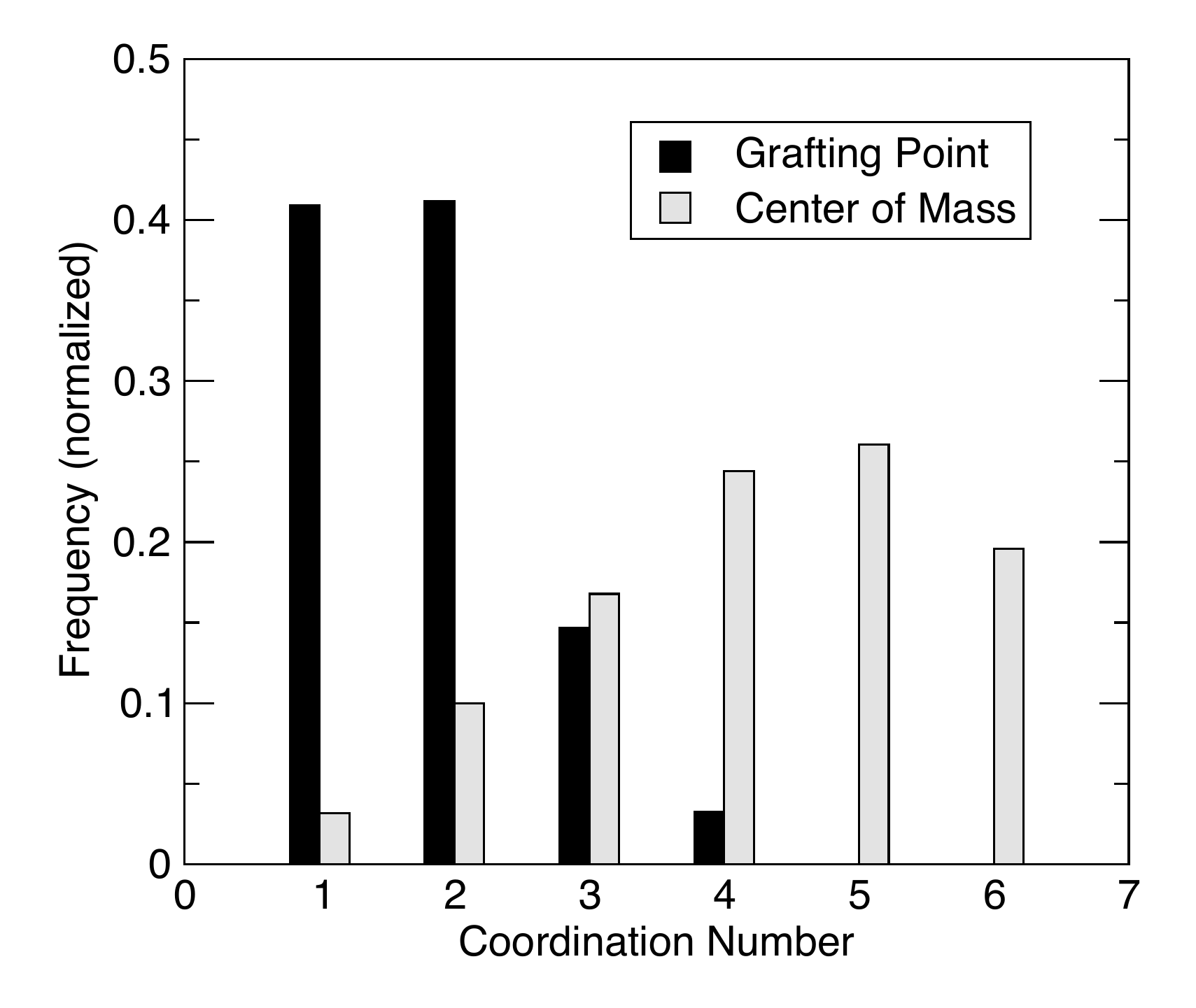} 
\caption{Histogram of the coordination number of centers-of-mass of the rods (grey) and a histogram of the coordination number of grafting points (black) for 1/\textit{T*} = 0.9.}
\label{figureCoordHistogram}
\end{figure}

\begin{figure}[h]
\includegraphics[width=3.5in]{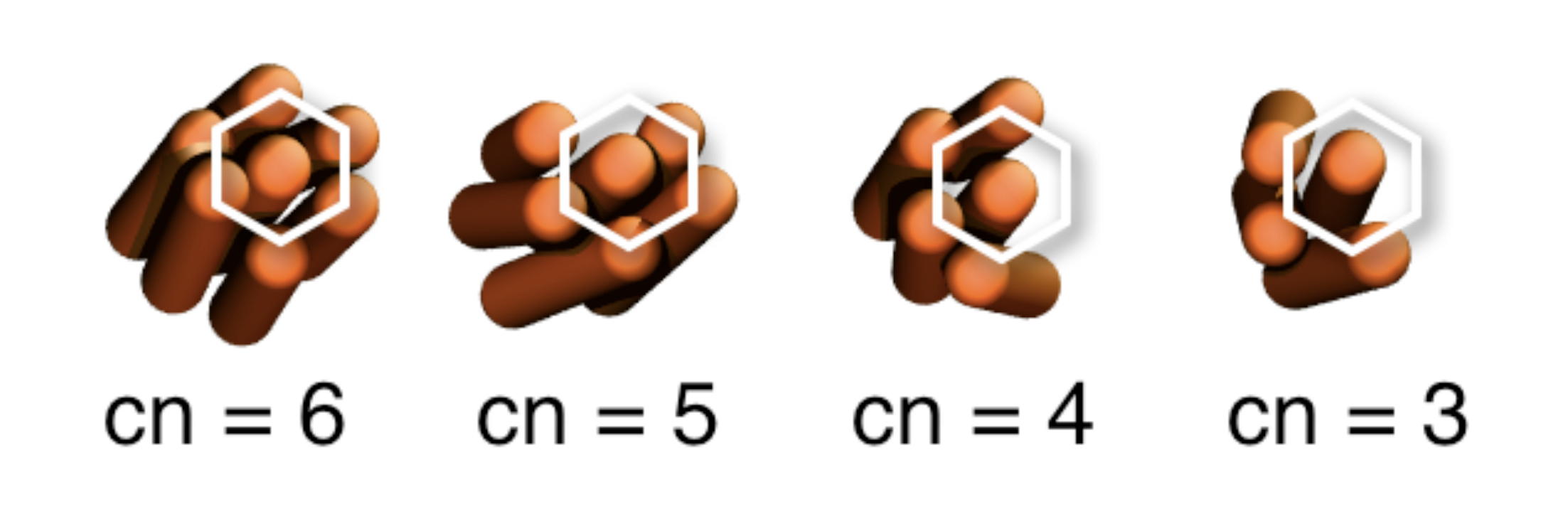} 
\caption{Example clusters formed by rods at 1/\textit{T*} = 0.9 for coordination numbers 6, 5, 4, and 3.}
\label{figureRodClusters}
\end{figure}

Similar to the TNS system, the DG formed by the TNR system also has distinct local ordering of the nanoparticles in addition to the bulk microphase separation.  Figure \ref{figureTNRNode} shows a simulation snapshot of the TNR node.  We notice that the nanorods attempt to adopt bundled structures, where a nanorod is surrounded by six nearest neighbors in a hexagonal fashion; a single bundle is highlighted in Figure \ref{figureTNRNode}.  A six neighbor bundle is the densest, minimal potential energy structure for seven rods, representing full coordination, analogous to a coordination of 12 for an icosahedron.  The tendency to form these bundles can be observed by examining the histogram of coordination number of the center-of-mass of each rod, shown in Figure \ref{figureCoordHistogram} for 1/\textit{T*} = 0.9.  There is a clear bias towards high coordination numbers and we find no coordinations greater than six.  We do find partially coordinated clusters as a result of rods being situated on the boundary with the tether region, as was also the case of the TNS system.  Example clusters from our simulations are shown in Figure \ref{figureRodClusters} for coordination numbers, cn = 3 - 6, where the preference to hexagonal packing is highlighted.

To reduce the grafting density of the tethers (i.e. the local density of tethers in a small region), thus maximizing entropy for the tether, we expect the grafting points of the tethers (i.e. the points in 3d space where the tether is attached to the nanorod) to be equally distributed along the interface between the nanorods and tethers.  In Figure \ref{figureCoordHistogram} we plot the histogram of the coordination number of the grafting points noticing a strong tendency for coordinations of one and two and do not find any coordinations above four.  A bundle of seven rods would have the most unfavorable configuration in terms of entropy if the grafting point coordination number were six, corresponding to all tethers being oriented on the same side of the bundle.  The histogram shows that each bundle has two or three tethers oriented in the same direction per bundle and thus tether attachment is well distributed.

We observe there is a deviation from the ideal hexagonally packed bundle structure as a result of the tether attachment; the bundles tilt and splay with respect to the central nanorod additionally reducing the grafting density of the tethers \cite{horsch2005}; this behavior is evident in the clusters in Figure \ref{figureRodClusters}.  This manifests itself as a twisted structure in the tubes (arms) of the DG, as has been seen in previous simulations of tethered nanorods that form twisted cylinders \cite{horsch2005}.  To quantify this behavior, we calculate the angle between the director of a nanorod with that of its nearest neighbors by calculating the dot product. On average we find that the angle between a nanorod and its neighbors is 10.7 degrees with a standard deviation of 6.1 degrees for a system with 1/\textit{T*} = 0.9.  Alternatively, we can quantify the splay of the rod bundles by calculating the nematic order parameter, \textit{S}, for each bundle.  To calculate the nematic order parameter, we first calculate the 3x3 nematic order tensor,
$$Q_{\alpha\beta} = \frac{1}{N}\sum_{i=1}^N\left(\frac{3}{2}u_{\alpha}^iu_{\beta}^i-\frac{\delta_{\alpha\beta}}{2}\right)$$ 
where $\alpha,\beta$ = x,y,z, $u_{\alpha}^i$ is the $\alpha$ component of the rod director and $\delta_{\alpha\beta}$ is the Kronecker delta \cite{dijkstra2003}.   We take \textit{S} to be the largest eigenvalue of the matrix Q \cite{dijkstra2003}.  This construction results in perfectly crystalline systems having a value of \textit{S}=1, nematic ordered liquid crystalline systems having \textit{S}=0.3-0.9, and isotropic systems having \textit{S}$<$0.3.  We find that for a system with 1/\textit{T*} = 0.9, the bundles of nanorods have an average value of \textit{S}= 0.96, demonstrating a small deviation from the ideal crystalline behavior.  Note that this is the average local nematic order parameter of the individual bundles, not the order parameter of the bulk system.  An alternative form for the nematic order parameter is given by $S=(1/2)(3cos^2\theta-1)$, where $\theta$ is the angle between a given nanrod and the director \cite{larsonbook}.  Substituting the average splay angle of 10.7 degrees into this form yields a value of \textit{S}=0.95 showing good agreement between these two methods of quantifying the splay.  

In Table \ref{tableTNRSplay} we calculate both the average splay angle and nematic order parameter of the bundles for an example cooling sequence, where the system is equilibrated at each 1/\textit{T*} prior to subsequent cooling; the transition from disordered to DG is denoted by the double line.  We notice that for the ordered microphases (1/\textit{T*} $>$  0.8) the splay angle is roughly 10 degrees, with a standard deviation of 7.2 degres for the lowest value of 1/\textit{T*}, and 3.8 degrees for the highest value of 1/\textit{T*}.  Within the ordered region, the nematic order parameter deviates little from the average value of 0.965.  Within the disordered region (1/\textit{T*} $\le$  0.8), both the splay angle and standard deviation are large, suggesting that there is a large variation in alignment of a set of neighboring rods; this is supported by values of \textit{S}  that are clearly less than the average value for the ordered DG configurations, indicating that neighboring rods are not strongly aligned. The combination of splay angle and \textit{S} allows us to conclude that, similar to the TNS system, there is a strong connection between microphase separation and local packing of the nanoparticles; we find strong local ordering where we find the DG, and very little local ordering elsewhere.

\begin{table}[h]
\caption{Local coordination in TNR system}
\begin{tabular}{cccc}
\hline
\;\; 1/\textit{T*}\;\; & \;\; $\chi$ $\pm$ error \;\; & \;\;  Splay Angle (degrees) $\pm$ stdev  \;\; & \;\;  S \\
\hline
0.50	& 2.17 $\pm$ 0.51 & 44.7 $\pm$ 25.7 & 0.73  \\
0.70	& 3.54 $\pm$ 0.61&  34.8 $\pm$ 25.7 & 0.79 \\
0.80	& 4.23 $\pm$ 0.66 & 19.6 $\pm$ 17.3 & 0.88\\
\hline\hline
0.85	& 4.58 $\pm$ 0.70 & 11.4 $\pm$ 7.2 & 0.96 \\
0.90	& 4.92 $\pm$ 0.72 & 10.7 $\pm$ 6.1 & 0.96 \\
0.95	& 5.27 $\pm$ 0.75 & 10.0 $\pm$ 4.7 & 0.97 \\
1.00	& 5.61 $\pm$ 0.77& 10.0 $\pm$ 4.7 & 0.97 \\
1.10	& 6.30 $\pm$ 0.83 & 9.7 $\pm$ 4.9 & 0.97 \\
1.20	& 6.99 $\pm$ 0.88 & 9.4 $\pm$ 3.8 &  0.97 \\
\hline
\end{tabular}
\label{tableTNRSplay}
\end{table}

\begin{figure}[h]
\includegraphics[width=5.5in]{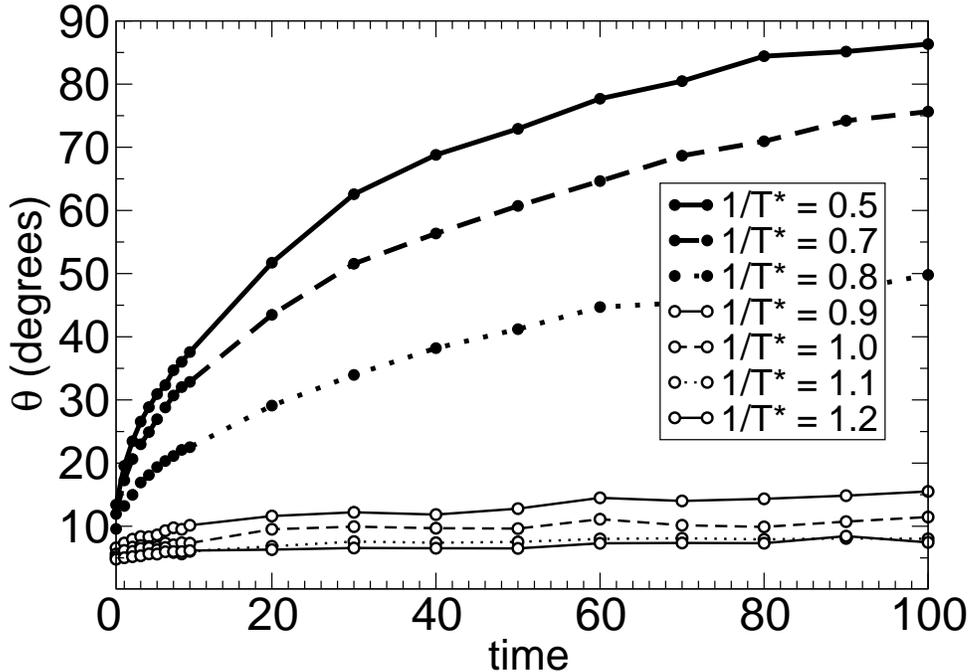} 
\caption{Average rotation as a function of time for various values of 1/\textit{T*}.  Solid black circles represent disordered microstructure, open circles correspond to the DG morphology.}
\label{figureRotationalBehavior}
\end{figure}

We also calculate the rotation of the rod as a function of time and 1/\textit{T*} to assess the ability of the rods to reorient themselves.  We calculate the average rotation of the rods by calculating the dot product of a rod's director at the current time with its initial starting director, as plotted in Figure \ref{figureRotationalBehavior}.  We find that at small values of 1/\textit{T*} there is a large amount of rotational freedom in the rods; we note that the rods for 1/\textit{T*}=0.5 have, on average, rotated approximately 85 degrees at time = 100 (10000 timesteps).  The large amount of rotational mobility explains the large splay angle and large standard deviation for high temperature disordered phases, as reported in Table \ref{tableTNRSplay}.  This rotational freedom is decreased as we increase 1/\textit{T*} (i.e. cool the system); it is clear that for values of  1/\textit{T*} $>$ 0.8, where the system has microphase separated into the DG, there is a drastic reduction in this rotational freedom and the average rotation of rods in the system is less than 15 degrees over the entire time range sampled.  The lack of rotational mobility suggests that the direction of the rod is strongly correlated through time and, again, this supports  the results in Table \ref{tableTNRSplay} where, for the DG, the distribution of splay angle is relatively narrow as compared to the disordered structure.

\section{\label{sec:conclusions}Conclusions}
We have performed Brownian dynamics simulations of tethered nanospheres and tethered nanorods that predict the double gyroid structure.   In both the tethered nanosphere and tethered nanorod systems we see a common trend towards the formation of dense, minimal potential energy structures.  Tethered nanospheres  predominantly form icosahedral clusters with partial coordination.  Tethered nanorods form hexagonally packed bundles of rods, with both full and partial coordination, where the rods tilt and splay with respect to their neighbors.  The icosahedral and hexagonal bundle structures adopted by the nanoparticles allow for a uniform density throughout the double gyroid structure; we do not see low density regions at the nodes of the double gyroid as would be expected for flexible chains  \cite{escobedo2005}.  The result of these local configurations is a reduction in packing frustration, which has been linked to the overall stability of the phase \cite{matsen1996,escobedo2005,escobedo2007}.  We have seen that the rigidity of the nanorods is crucial for the stabilization of the double gyroid relative to BCPs; for equivalent conditions, we did not find the double gyroid structure for an analogous flexible diblock copolymer system.  This suggests that by properly choosing the geometry of the nanoparticle, we may be able to not only stabilize the double gyroid, but potentially other bicontinuous structures by harnessing the unique combination of microphase separation and local packing.

\section{Acknowledgements}
We thank A.S. Keys, R.G. Larson, and Z-L. Zhang for useful discussions and  the Department of Energy (DE-FG02-02ER46000) and Department of Education (GAANN) for funding.

\bibliographystyle{ieeetr}
\bibliography{Iacovella_gyroid}

\begin{thebibliography}{10}

\bibitem{bates1990}
F.~S. Bates and G.~Fredrickson, ``Block copolymer thermodynamics - theory and
  experiment,'' {\em Annual Review of Physical Chemistry}, vol.~41,
  pp.~525--557, 1990.

\bibitem{larson1996}
R.~G. Larson, ``Monte carlo simulations of the phase behavior of surfactant
  solutions,'' {\em Journal De Physique II}, vol.~6, no.~10, pp.~1441--1463,
  1996.

\bibitem{maldovan2002}
M.~Maldovan, A.~M. Urbas, N.~Yufa, W.~C. Carter, and E.~L. Thomas, ``Photonic
  properties of bicontinuous cubic microphases,'' {\em Physical Review B},
  vol.~65, no.~16, 2002.
\newblock 165123.

\bibitem{pine2005}
J.~Chiu, B.~Kim, E.~Kramer, and D.~Pine, ``Control of nanoparticle location in
  block copolymers,'' {\em Journal of the American Chemical Society}, vol.~127,
  pp.~5036--5037, April 2005.

\bibitem{kotov2002}
S.~Westenhoff and N.~A. Kotov, ``Quantum dot on a rope,'' {\em Journal of the
  American Chemical Society}, vol.~124, no.~11, pp.~2448--2449, 2002.

\bibitem{song2003}
T.~Song, S.~Dai, K.~C. Tam, S.~Y. Lee, and S.~H. Goh, ``Aggregation behavior of
  c-60-end-capped poly(ethylene oxide)s,'' {\em Langmuir}, vol.~19, no.~11,
  pp.~4798--4803, 2003.

\bibitem{frank2005}
Y.~Kim, J.~Pyun, J.~M.~J. Frechet, C.~J. Hawker, and C.~W. Frank, ``The
  dramatic effect of architecture on the self-assembly of block copolymers at
  interfaces,'' {\em Langmuir}, vol.~21, no.~23, pp.~10444--10458, 2005.

\bibitem{nie2007}
Z.~Nie, D.~Fava, E.~Kumacheva, S.~Zou, G.~Walker, and M.~Rubinstein,
  ``Self-assembly of metal-polymer analogues of amphiphilic triblock
  copolymers,'' {\em Nature Materials}, vol.~6, pp.~609--614, August 2007.

\bibitem{devries2007}
G.~DeVries, M.~Brunnbauer, Y.~Hu, A.~Jackson, B.~Long, B.~Neltner, O.~Uzun,
  B.~Wunsch, and F.~Stellacci, ``Divalent metal nanoparticles,'' {\em Science},
  vol.~315, pp.~358--361, January 2007.

\bibitem{date2003}
R.~Date and D.~Bruce, ``Shape amphiphiles: Mixing rods and disks in liquid
  crystals,'' {\em Journal of the American Chemical Society}, vol.~125, no.~30,
  pp.~9012--9013, 2003.

\bibitem{zhang2003}
Z.~L. Zhang, M.~A. Horsch, M.~H. Lamm, and S.~C. Glotzer, ``Tethered nano
  building blocks: Toward a conceptual framework for nanoparticle
  self-assembly,'' {\em Nano Letters}, vol.~3, no.~10, pp.~1341--1346, 2003.

\bibitem{horsch2005}
M.~A. Horsch, Z.~L. Zhang, and S.~C. Glotzer, ``Self-assembly of
  polymer-tethered nanorods,'' {\em Physical Review Letters}, vol.~95, no.~5,
  2005.
\newblock 056105.

\bibitem{iacovella2005}
C.~R. Iacovella, M.~A. Horsch, Z.~Zhang, and S.~C. Glotzer, ``Phase diagrams of
  self-assembled mono-tethered nanospheres from molecular simulation and
  comparison to surfactants,'' {\em Langmuir}, vol.~21, no.~21, pp.~9488--9494,
  2005.

\bibitem{iacovella2007}
C.~R. Iacovella, A.~S. Keys, M.~A. Horsch, and S.~C. Glotzer, ``Icosahedral
  packing of polymer-tethered nanospheres and stabilization of the gyroid
  phase,'' {\em Physical Review E}, vol.~75, April 2007.

\bibitem{horsch2006}
M.~A. Horsch, Z.~L. Zhang, and S.~C. Glotzer, ``Simulation studies of
  self-assembly of end-tethered nanorods in solution and role of rod aspect
  ratio and tether length,'' {\em Journal of Chemical Physics}, vol.~125,
  November 2006.

\bibitem{lee1998}
M.~Lee, B.~Cho, H.~Kim, J.~Yoon, and W.~Zin, ``Self-organization of rod-coil
  molecules with layered crystalline states into thermotropic liquid
  crystalline assemblies,'' {\em Journal of the American Chemical Society},
  vol.~120, pp.~9168--9179, September 1998.

\bibitem{cho2004}
B.~K. Cho, A.~Jain, S.~M. Gruner, and U.~Wiesner, ``Mesophase
  structure-mechanical and ionic transport correlations in extended amphiphilic
  dendrons,'' {\em Science}, vol.~305, no.~5690, pp.~1598--1601, 2004.

\bibitem{soddemann2001}
T.~Soddemann, B.~Dunweg, and K.~Kremer, ``A generic computer model for
  amphiphilic systems,'' {\em European Physical Journal E}, vol.~6,
  pp.~409--419, December 2001.

\bibitem{grest1986}
G.~S. Grest and K.~Kremer, ``Molecular dynamics simulation for polymers in the
  presence of a heat bath,'' {\em Phys. Rev. A.}, vol.~33, pp.~3628--3631,
  1986.

\bibitem{solomon2007}
K.~E. Sung, S.~A. Vanapalli, D.~Mukhija, H.~A. McKay, J.~M. Millunchick, M.~A.
  Burns, and M.~J. Solomon, ``Programmable fluidic synthesis of microparticles
  with configurable anisotropy,'' {\em Journal of the American Chemical
  Society}, 2007.

\bibitem{siva2007}
S.~A. Vanapalli, C.~R. Iacovella, K.~E. Sung, D.~Mukhija, H.~A. McKay, J.~M.
  Millunchick, M.~A. Burns, S.~C. Glotzer, and M.~J. Solomon, ``Fluidic
  assembly and packing of microspheres in confined channels,'' {\em Langmuir},
  vol.~24, no.~7, pp.~3661--3670, 2008.

\bibitem{horsch2004}
M.~Horsch, Z.-L. Zhang, C.~Iacovella, and S.~Glotzer, ``Hydrodynamics and
  microphase ordering in block copolymers: Are hydrodynamics required for
  ordered phases with periodicity in more than one dimension?,'' {\em Journal
  of Chemical Physics}, vol.~121, pp.~11455--11462, December 2004.

\bibitem{allentildesley}
M.~P. Allen and D.~Tildesley, {\em Computer Simulations of Liquids}.
\newblock Oxford University Press, 1st~ed., 1987.

\bibitem{groot1999}
R.~D. Groot, T.~J. Madden, and D.~J. Tildesley, ``On the role of hydrodynamic
  interactions in block copolymer microphase separation,'' {\em Journal of
  Chemical Physics}, vol.~110, pp.~9739--9749, May 1999.

\bibitem{larsonbook}
R.~Larson, {\em The Structure and Rheology of Complex Fluids}.
\newblock Oxford University Press, 1998.

\bibitem{hajduk1994}
D.~A. Hajduk, P.~E. Harper, S.~M. Gruner, C.~C. Honeker, G.~Kim, E.~L. Thomas,
  and L.~J. Fetters, ``The gyroid - a new equilibrium morphology in weakly
  segregated diblock copolymers,'' {\em Macromolecules}, vol.~27, no.~15,
  pp.~4063--4075, 1994.

\bibitem{arthi2008}
A.~Jayaraman and K.~Schweizer, ``Structure and assembly of dense solutions and
  melts of single tethered nanoparticles,'' {\em Journal of Chemical Physics},
  vol.~128, no.~16, p.~164904, 2008.

\bibitem{bates1996}
M.~W. Matsen and F.~S. Bates, ``Unifying weak- and strong-segregation block
  copolymer theories,'' {\em Macromolecules}, vol.~29, no.~4, pp.~1091--1098,
  1996.

\bibitem{escobedo2006}
F.~J. Martinez-Veracoechea and F.~A. Escobedo, ``Simulation of the gyroid phase
  in off-lattice models of pure diblock copolymer melts,'' {\em Journal of
  Chemical Physics}, vol.~125, no.~10, 2006.
\newblock 104907.

\bibitem{rychkov2005}
I.~Rychkov, ``Block copolymers under shear flow,'' {\em Macromolecular Theory
  and Simulations}, vol.~14, no.~4, pp.~207--242, 2005.

\bibitem{cochran2006}
E.~Cochran, C.~Garcia-Cervera, and G.~Fredrickson, ``Stability of the gyroid
  phase in diblock copolymers at strong segregation,'' {\em Macromolecules},
  vol.~39, pp.~2449--2451, April 2006.

\bibitem{lee2001}
M.~Lee and W.~Cho, B.K.~Zin, ``Supramolecular structures from rod-coil block
  copolymers,'' {\em Chemical Reviews}, vol.~101, pp.~3869--3892, December
  2001.

\bibitem{matsen1996}
M.~W. Matsen and F.~S. Bates, ``Origins of complex self-assembly in block
  copolymers,'' {\em Macromolecules}, vol.~29, no.~23, pp.~7641--7644, 1996.

\bibitem{hasegawa1996}
H.~Hasegawa, T.~Hashimoto, and S.~T. Hyde, ``Microdomain structures with
  hyperbolic interfaces in block and graft copolymer systems,'' {\em Polymer},
  vol.~37, no.~17, pp.~3825--3833, 1996.

\bibitem{escobedo2005}
F.~J. Martinez-Veracoechea and F.~A. Escobedo, ``Lattice monte carlo
  simulations of the gyroid phase in monodisperse and bidisperse block
  copolymer systems,'' {\em Macromolecules}, vol.~38, no.~20, pp.~8522--8531,
  2005.

\bibitem{escobedo2007}
F.~J. Martinez-Veracoechea and F.~A. Escobedo, ``Monte carlo study of the
  stabilization of complex bicontinuous phases in diblock copolymer systems,''
  {\em Macromolecules}, vol.~40, pp.~7354--7365, October 2007.

\bibitem{steinhardt1983}
P.~J. Steinhardt, D.~R. Nelson, and M.~Ronchetti, ``Bond-orientational order in
  liquids and glasses,'' {\em Physical Review B}, vol.~28, no.~2, pp.~784--805,
  1983.

\bibitem{keys2008}
A.~S. Keys, C.~R. Iacovella, and S.~C. Glotzer {\em In preparation}, 2008.

\bibitem{doye1996}
J.~P.~K. Doye and D.~J. Wales, ``The effect of the range of the potential on
  the structure and stability of simple liquids: from clusters to bulk, from
  sodium to c60,'' {\em Journal of Physics B: Atomic, Molecular and Optical
  Physics}, vol.~29, no.~21, pp.~4859--4894, 1996.

\bibitem{doye2001}
J.~P.~K. Doye, D.~J. Wales, and S.~I. Simdyankin, ``Global optimization and the
  energy landscapes of dzugutov clusters,'' {\em Faraday Discussions},
  vol.~118, pp.~159--170, 2001.

\bibitem{dijkstra2003}
M.~Cosentino~Lagomarsino, M.~Dogterom, and M.~Dijkstra, ``Isotropic-nematic
  transition of long, thin, hard spherocylinders confined in a
  quasi-two-dimensional planar geometry,'' {\em Journal of Chemical Physics},
  vol.~119, pp.~3535--3540, August 2003.

\end{thebibliography}

\end{document}